\documentclass[prl,preprint,superscriptaddress,showpacs,amsmath,amssymb,aps]{revtex4-1}

\usepackage{graphicx}
\usepackage{amsmath}
\usepackage[version=3]{mhchem}
\usepackage[usenames]{color}
\usepackage{color}
\usepackage{physics}
\usepackage[colorlinks=true, linkcolor=blue, citecolor=blue, urlcolor=blue,pdftex]{hyperref}
\definecolor{newcolor}{rgb}{0.9,0,0.1}

\usepackage{orcidlink}
\usepackage{multirow}
\usepackage{siunitx}
\usepackage{float}

\newcommand{\figref}[1]{Fig.~\ref{#1}}

\usepackage{xr}
\makeatletter
\newcommand*{\addFileDependency}[1]{
\typeout{(#1)}
\@addtofilelist{#1}
\IfFileExists{#1}{}{\typeout{No file #1.}}

\makeatother
}
\newcommand*{\myexternaldocument}[1]{%
\externaldocument{#1}%
\addFileDependency{#1.tex}%
\addFileDependency{#1.aux}%
}
\myexternaldocument{supplement}

\begin{document}

\title{Layer-Dependent Charge State Lifetime of Single Se Vacancies in \ce{WSe2}}

\author{Laric Bobzien\,\orcidlink{0000-0000-0000-0000} \textsuperscript{\textdagger}}

\affiliation{nanotech@surfaces Laboratory, Empa -- Swiss Federal Laboratories for Materials Science and Technology, D\"ubendorf 8600, Switzerland}

\author{Jonas Allerbeck\,\orcidlink{0000-0002-3912-3265} \textsuperscript{\textdagger}}
\affiliation{nanotech@surfaces Laboratory, Empa -- Swiss Federal Laboratories for Materials Science and Technology, D\"ubendorf 8600, Switzerland}

\author{Nils Krane\, \orcidlink{0009-0004-6970-0846}}
\affiliation{nanotech@surfaces Laboratory, Empa -- Swiss Federal Laboratories for Materials Science and Technology, D\"ubendorf 8600, Switzerland}

\author{Andres Ortega-Guerrero\,\orcidlink{0000-0002-0065-0623}}
\affiliation{nanotech@surfaces Laboratory, Empa -- Swiss Federal Laboratories for Materials Science and Technology, D\"ubendorf 8600, Switzerland}

\author{Zihao Wang}
\affiliation{Department of Materials Science and Engineering, The Pennsylvania State University, University Park, PA 16082, USA}

\author{Daniel E. Cintron Figueroa}
\affiliation{Department of Materials Science and Engineering, The Pennsylvania State University, University Park, PA 16082, USA}

\author{Chengye Dong}
\affiliation{Two-Dimensional Crystal Consortium, The Pennsylvania State University, University Park, PA 16802, USA}


\author{Carlo A. Pignedoli\,\orcidlink{0000-0002-8273-6390}}
\affiliation{nanotech@surfaces Laboratory, Empa -- Swiss Federal Laboratories for Materials Science and Technology, D\"ubendorf 8600, Switzerland}

\author{Joshua A. Robinson\,\orcidlink{0000-0002-1513-7187}}
\affiliation{Department of Materials Science and Engineering, The Pennsylvania State University, University Park, PA 16082, USA}
\affiliation{Two-Dimensional Crystal Consortium, The Pennsylvania State University, University Park, PA 16802, USA}
\affiliation{Department of Chemistry and Department of Physics, The Pennsylvania State University, University Park, PA, 16802, USA}

\author{Bruno Schuler\,\orcidlink{0000-0002-9641-0340}}
\email[]{bruno.schuler@empa.ch}
\affiliation{nanotech@surfaces Laboratory, Empa -- Swiss Federal Laboratories for Materials Science and Technology, D\"ubendorf 8600, Switzerland}

\begin{abstract}
Defect engineering in two-dimensional semiconductors has been exploited to tune the optoelectronic properties and introduce new quantum states in the band gap. Chalcogen vacancies in transition metal dichalcogenides in particular have been found to strongly impact charge carrier concentration and mobility in 2D transistors as well as feature sub-gap emission and single-photon response.
In this letter, we investigate the layer-dependent charge state lifetime of Se vacancies in WSe$_2$. In one monolayer WSe$_2$, we observe ultrafast charge transfer from the lowest unoccupied orbital of the top Se vacancy to the graphene substrate within $(1\pm0.2)$\,ps measured via the current saturation in scanning tunneling approach curves. For Se vacancies decoupled by TMD multilayers, we find a sub-exponential increase of the charge lifetime from $(62\pm 14)$\,ps in bilayer to few nanoseconds in four-layer WSe\textsubscript{2}, alongside a reduction of the defect state binding energy. Additionally, we attribute the continuous suppression and energy shift of the d$I$/d$V$ in-gap defect state resonances at very close tip--sample distances to a current saturation effect. Our results provide a key measure of the layer-dependent charge transfer rate of chalcogen vacancies in TMDs.
\end{abstract}

\date{\today}
\maketitle


\section{Introduction}
Deep-level traps in semiconductors are typically undesired crystal imperfections with binding energies much larger than the thermal energy, leading to electron trapping and persistent photoconductivity~\cite{mooney_Deep_1990,queisser_Defects_1998}. Transient spectroscopy and temperature-dependent Hall-effect measurements have helped to elucidate the binding energy and capture-release mechanisms in the elusive DX center in III-V semiconductors~\cite{theis_DX_1990, criado_Deep_1986, oelgart_Hall_1990}, however, microscopic identification has posed a longstanding challenge. In photonics, deep levels within wide-band semiconductors constitute artificial atoms with appealing properties such as narrow sub-gap emission~\cite{sedhain_Nature_2010, suttrop_Boronrelated_1990} and single-photon response~\cite{zhang_Material_2020,aharonovich_Solidstate_2016}. Spin-bearing deep defects hold great potential for quantum computing or communication protocols~\cite{weber_Quantum_2010,awschalom_Quantum_2013,wolfowicz_Quantum_2021}. Thereby, the charge state decisively impacts the optical properties and magnetic moment of deep centers.\\

In two-dimensional (2D) semiconductors, deep-level defects are prevalent due to reduced screening favoring large binding energies~\cite{zhu_DimensionalityInhibited_2021}. However, tunneling barriers tend to be small as heterostructures typically consist of few layers only, facilitating the release of captured charge carriers by resonant tunneling~\cite{mishchenko_Twistcontrolled_2014}. For both the defect binding energy and the charge state lifetime, the coupling between layers is decisive. Semiconductor-metal contacts formed by transition metal dichalcogenide (TMD) and graphene are among the most prevalent material combinations used for fabricating 2D vertical field-effect transistors~\cite{chhowalla_Twodimensional_2016,roy_DualGated_2015} or light emitting diodes~\cite{withers_Lightemitting_2015,withers_WSe2_2015}. Understanding the inter-layer charge dynamics in such van der Waals (vdW) heterostructures is essential for improving their performance. Time- and angle-resolved photoemission spectroscopy (tr-ARPES) measurements
revealed ultrafast inter-layer charge transfer in epitaxial WS$_2$/graphene heterostructures following photoexcitation~\cite{aeschlimann_Direct_2020}, suggesting momentum- and energy-matched inter-layer charge transfer as well as defect-assisted tunneling as the main mechanisms~\cite{krause_Microscopic_2021}.\\

Here, we determine the layer-dependent binding energy and charge state lifetime at a single chalcogen vacancy, the most iconic deep-level defect in TMDs, supported by an epitaxial graphene layer. By means of low-temperature scanning tunneling spectroscopy (STS) we directly measure the in-gap defect states of single Se vacancies (Vac$_\text{Se}$) within the quasiparticle band gap of \ce{WSe2} as a function of layer thickness. The average charge residence time of electrons within specific defect states could be determined by saturation of the tunneling current in scanning tunneling microscopy (STM) approach curves~\cite{steurer_Local_2014,kaiser_Chargestate_2023}. 
By increasing the layer count from monolayer \ce{WSe2} to four layer \ce{WSe2} we observe the transition from a deep level to a shallow defect state, while counterintuitively the charge lifetime increases from 1\,ps to 4\,ns.
The microscopic picture of transient charging of single deep levels in TMDs provides insights into charge relaxation mechanisms and the degree of decoupling in multi-layer vdW heterostructures.

\section{Electronic structure of Se vacancies in mono- and multilayer WSe$_2$}

\begin{figure}[H]
\centering
\includegraphics[width=0.5\textwidth]{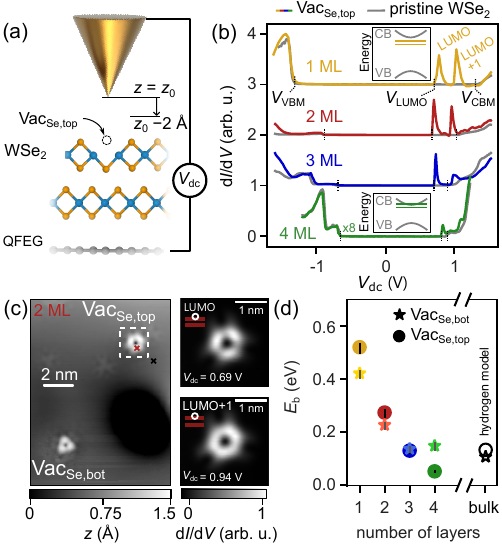}
\caption{\label{fig:fig1}
Single selenium vacancies in few-layer WSe\textsubscript{2}. (a) Schematic illustration of a top Se vacancy (Vac\textsubscript{Se,top}) in 2\,ML WSe\textsubscript{2} on quasi free-standing epitaxial graphene (QFEG) in an STM tunnel junction. The tip--sample distance $z$, tunneling set point $z_0$, and bias voltage $V_\mathrm{dc}$ are indicated. (b) Differential conductance measurements (d$I$/d$V$) recorded on Vac\textsubscript{Se,top} in the uppermost layer of 1-4\,ML WSe\textsubscript{2}. Dotted ticks mark the onset voltages for the lowest in-gap defect state LUMO, as well as for valence band maxima (VBM) and conduction band minima (CBM), respectively. Grey curves reference the pristine WSe\textsubscript{2} surface.
(c) Overview image of a WSe\textsubscript{2} bilayer with a top and bottom Se vacancy ($V_\mathrm{dc}=1.2$\,V, $I_0=100$\,pA). Right panels: Constant height d$I$/d$V$ maps of the Vac\textsubscript{Se,top} defect orbitals. (d) Binding energy ($E_\mathrm{b}=E_\mathrm{LUMO}-E_\mathrm{CBM}$) of the LUMO state, determined from spectra in (b) and Fig.~S1, Fig.~S2. Binding energy values include a correction of the effective tip--sample voltage, where error bars reflect an uncertainty of the absolute distance (Supplementary Information). The binding energies  from the 1\,ML case are extrapolated with an effective hydrogen model to obtain the expected bulk value.}
\end{figure}

WSe$_2$ few-layer samples are grown via metal organic chemical vapor deposition on hydrogen intercalated quasi-freestanding graphene on doped silicon carbide~\cite{eichfeld_Highly_2015}. Isolated Se vacancies are created by light sputtering of the sample with argon ions (1\,s, $5.2\cdot10^{-6}$\,mbar, 0.2\,kV bias, sub-\,$\mu$A ion current on sample)~\cite{trishin_Electronic_2023}. \figref{fig:fig1}(a) shows a schematic cross section of the experiment. 
STM/STS measurements are performed at $<5$\,K base temperature with an Au coated tungsten tip.
We identify top and bottom Se vacancies by their characteristic STS fingerprint [\figref{fig:fig1}(b)] and orbital shape~\cite{schuler_Large_2019}, considering only isolated defects with a few nanometer lateral separation from other defects [\figref{fig:fig1}(c)]. Right panels in \figref{fig:fig1}(c) depict the constant-height orbital distribution of the lowest unoccupied defect states denoted LUMO and LUMO+1 labelled in (b), supported by density functional theory (DFT) calculations [Fig.~S4]. The reference tip--sample distance $z_0$ is defined from a tunneling set point approximately 200\,mV above the onset of the conduction band of \ce{WSe2}, corresponding to ($V_0=1.5$\,V, $I_0=100$\,pA) for 1\,ML and ($V_0=1.2$\,V, $I_0=100$\,pA) for 2-4\,ML WSe\textsubscript{2}. 

As reported previously~\cite{stolz_Layerdependent_2022}, additional steps in the local density of states of the valence band provide a simple means to unambiguously identify the number of layers. As expected, the band gap reduces for increasing number of layers\cite{stolz_Layerdependent_2022} [\figref{fig:fig1}(b), Fig.~S8] concurrent with a strong reduction of the defect binding energy from 594\,mV for 1\,ML to 351\,mV for 2\,ML (186\,mV for 3\,ML) down to 82\,mV for 4\,ML [\figref{fig:fig1}(d)], marking the transition from a deep level state to a shallow defect.
Similar trends have been observed for layer-dependent exciton binding energies~\cite{chernikov_Exciton_2014}.
We estimate the binding energy of the Vac\textsubscript{Se} in the bulk by extrapolating the binding energy of the single monolayer via an effective hydrogen model~\cite{ihn_Semiconductor_2009}. The difference between the prediction of this simplified model and the experimentally observed trend hints towards delocalization and screening effects that impact the electronic properties of the vacancy.

\section{Charge state lifetime via saturation of tunneling rate}
\begin{figure}[H]
\centering
\includegraphics[width=\textwidth]{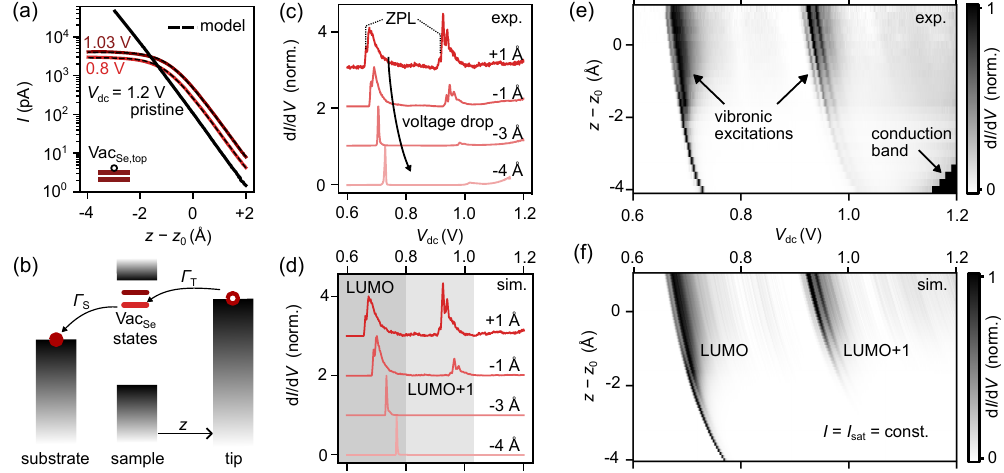}
\caption{\label{fig:fig2}
STM approach curves and d$I$/d$V$ spectra of a Se vacancy in bilayer WSe\textsubscript{2} at reduced tip--sample distances. (a) STM approach curves ($z$ spectroscopy) of the LUMO and LUMO+1 orbital states and an unperturbed WSe\textsubscript{2} reference (black). $z_0 = 0$\,Å is defined by a set point of (1.2\,V, 100\,pA) on a pristine region of the sample. Dashed lines are fits of the double barrier model in Eq.~\eqref{eq:DoubleBarrier}, with a charge state lifetime of 55.9\,ps and $\kappa$ = 1.06\,Å$^{-1}$ for the LUMO and 40.4\,ps, $\kappa$ = 1.05\,Å$^{-1}$ for the LUMO+1. (b) Level diagram of a double barrier tunneling junction with charge transfer rates $\Gamma_\mathrm{T}$ and $\Gamma_\mathrm{S}$. (c,d) Measured and calculated d$I$/d$V$ spectra as a function of tip--sample distances measured at the Vac\textsubscript{Se}. Individual spectra are normalized to unity at the LUMO resonance. (e,f) d$I$/d$V$ spectra as a function of tip-sample distance, expanding the data in panels (c,d). The calculation of $z$-dependent d$I$/d$V$ is based on the d$I$/d$V$ spectrum at $z_\mathrm{0}$+1\,Å in (c) scaled with the charge state lifetime and $\kappa$ for LUMO (55.9\,ps, $\kappa$ = 1.06\,Å$^{-1}$) as obtained through fitting (a). We estimate the voltage drop across the WSe\textsubscript{2} using a plate capacitor model, assuming the sample surface at $z_0-9$\,Å, a thickness of 6.5\,Å \cite{lin_Atomically_2014} per monolayer WSe\textsubscript{2}, and a dielectric constant of 6.5 \cite{hou_Quantification_2022}.}
\end{figure}

In the following, we exemplify our approach to measure the charge state lifetime of a single defect on the Se top vacancy (Vac\textsubscript{Se,top}) in the upper layer of 2\,ML and 3\,ML WSe\textsubscript{2}, respectively. By reducing the tip--sample distance on a pristine region of the 2\,ML WSe\textsubscript{2} sample we find the expected exponential increase of the tunneling current $I(z)=I_0\exp{-2\kappa z}$ with $\kappa=1.06\,\text{\AA}^{-1}$, indicated by the black curve in \figref{fig:fig2}(a). When tunneling to a localized in-gap state of a Vac\textsubscript{Se}, this exponential scaling saturates, limited by the maximum tunneling rate $\Gamma_\mathrm{S}$ from the defect orbital to the graphene substrate due to Coulomb blocking of this tunneling channel [\figref{fig:fig2}(b)]. The charge state lifetime is given by the inverse tunneling rate from the defect to the graphene ($\tau=\Gamma_\mathrm{S}^{-1}$). We extract tip and substrate tunneling rates ($\Gamma_\mathrm{T}$ and $\Gamma_\mathrm{S}$) and the exponential scaling $\kappa$ by fitting the saturated approach curves with a double barrier sequential tunneling junction model described by the following rate equation
\begin{equation}\label{eq:DoubleBarrier}
    I(z) = \frac{e}{\Gamma_\mathrm{S}^{-1} + \Gamma_\mathrm{T}^{-1}e^{2z\kappa}}. 
\end{equation}

From the saturation current, we extract a charge state lifetime of 55.9\,ps for this Vac$_\text{Se}$ LUMO in 2\,ML \ce{WSe2}. At 248\,meV higher energy, above the LUMO+1 resonance, we measure an effective charge state lifetime of 40.4\,ps. While the LUMO is an isolated independent orbital state, voltages resonant with the LUMO+1 inevitably enable tunneling through the LUMO state, which implies that the associated tunneling rate is an effective mixture of both orbital states. We disentangle individual contributions by estimating the probability of tunneling via either state based on the reduced potential barrier for the elevated orbital state and the wavefunction overlap in DFT calculations (Supplementary Information). As result, we obtain a pure LUMO+1 charge state lifetime of 35\,ps, where the elevated energy causes approximately 29\,\% reduction of the LUMO+1 charge state lifetime compared to the LUMO. Based on this, we estimate that the LUMO+1 orbital delocalization contributes additional 15\,\% reduction.

Concurrent with this saturation, we observe a continuous suppression and energy shift of the d$I$/d$V$ spectra of the LUMO and LUMO+1 resonances as the tip--sample distance $z$ decreases [\figref{fig:fig2}(c,e)]. This behavior is well reproduced by calculations based on the rate equation model [\figref{fig:fig2}(d,f)], and further detailed in the Supplementary Information. The consistent electrostatic shift of the LUMO onset towards higher bias voltage aligns with expectations of a double barrier tunneling junction geometry, arising from a larger proportion of the voltage drop occurring across the TMD layer(s), effectively reducing the vacuum gap voltage at reduced $z$~\cite{nazin_Tunneling_2005}. 
The suppression of LUMO+1 and spectral narrowing of the LUMO state in d$I$d$V$ is a consequence of the saturated tunneling current, where in saturation a bias modulation for lock-in detection does not modulate the tunnel current any longer, hence suppressing the signal. In other words, a partial region of the local density of states (LDOS) at the lowest energy is sufficient to reach the saturation current, and due to Coulomb blocking localized higher energy states become unavailable when the defect orbital is dynamically charged. Consequently, only d$I$/d$V$ spectra measured in the exponential regime of the approach curve can be correlated with the sample LDOS. Unsaturated spectra reveal the zero-phonon line (ZPL) and vibronic excitations of LUMO and LUMO+1 states that vanish as the current saturates already at the ZPL.

\begin{figure}[H]
\centering
\includegraphics[width=0.5\textwidth]{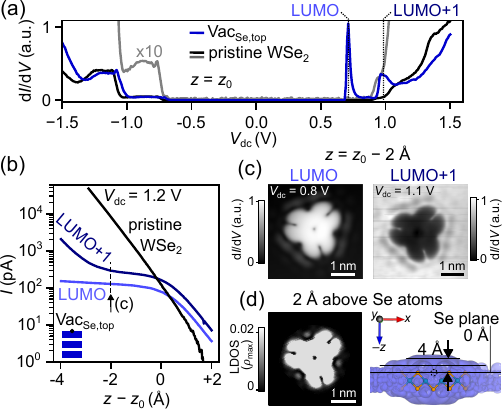}
\caption{\label{fig:fig3}
Orbital mapping of Se vacancy states in trilayer WSe$_2$ in the saturation regime. (a) d$I$/d$V$ spectrum of Vac\textsubscript{Se,top} (blue) and pristine 3\,ML WSe\textsubscript{2} (black) emphasizing an overlap of LUMO+1 and the conduction band (grey). (b) Approach curves measured for LUMO and LUMO+1 resonances at the Vac\textsubscript{Se,top} and for the CB of pristine WSe\textsubscript{2}. (c) Constant height d$I$/d$V$ maps of the Vac\textsubscript{Se,top} LUMO and LUMO+1 at -2\,Å in the current saturation regime. (d) DFT calculations of the LUMO electron density isosurface at $\rho_\mathrm{max}=6 \cdot 10^{-7}$\, $e/a_\mathrm{0}^3$ ($a_\mathrm{0}$ the Bohr radius) (right) and a constant height electron density image at 2\,Å above the Se atoms (left).}
\end{figure}

In \figref{fig:fig3}, we apply the same experimental approach to examine a Vac\textsubscript{Se,top} in 3\,ML WSe\textsubscript{2}. The saturation current for the LUMO resonance is reduced by more than one order of magnitude corresponding to an increased charge residence time of 1.09\,ns due to stronger decoupling from the substrate. Each layer of WSe\textsubscript{2} increases the distance between vacancy and substrate by approximately 6.5\,\si{\angstrom} ~\cite{lin_Atomically_2014}, reducing the charge transfer rate to the sample while the orbital overlap of Vac\textsubscript{Se} and STM tip remains similar. 
The most significant difference compared to 2\,ML is the current increase past the saturation value at close distances for voltages resonant with the LUMO+1 state. This behavior is due to the resonant character of the LUMO+1 state in 3\,ML \ce{WSe2}, which is located above the CBM. Hence, tunneling into dispersive CB states is possible despite dynamic charging of the defect, causing a further increase of the tunneling current. As the LUMO is strictly bound in the band gap, no increase is observed there.

At reduced tip--sample distances where the current saturates ($z\leq z_0-2\si{\angstrom}$), the orbital maps exhibit a fundamentally different electronic contrast compared to those in the exponential tunneling regime [cf. \figref{fig:fig1}(c)]. Notably, the contrast in the periphery of the defect state wavefunction is enhanced, while the main lobes with the highest LDOS become saturated as seen in \figref{fig:fig3}(c). We attribute the observed contrast inversion for the LUMO+1 at $V=1.1$\,V to predominant tunneling into the conduction band, which is unaffected by saturation effects. At this voltage, the populated defect state causes a transient band bending that manifests as current suppression and resembles the defect orbital.

\section{Layer-dependent variation of the charge transfer time}

\begin{figure}[H]
\centering
\includegraphics[width=0.5\textwidth]{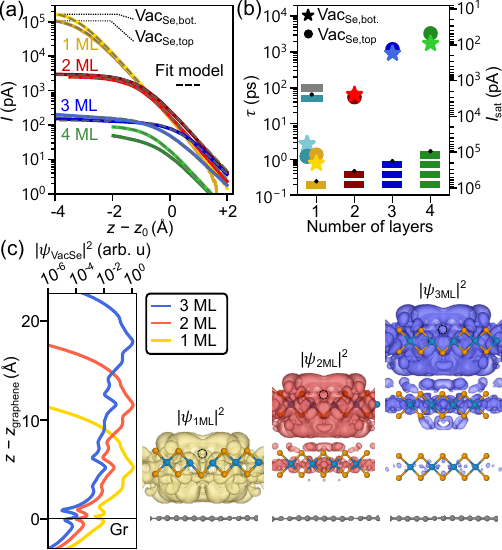}
\caption{\label{fig:fig4}
Layer dependence of the charge state lifetime of Se vacancies. (a) STM approach curves at the LUMO resonance of Vac\textsubscript{Se} in the uppermost layer of 1-4\,ML WSe\textsubscript{2} for bottom (light color) and top (dark color) Se vacancies with model fits (dashed lines). (b) Layer-dependent charge state lifetime for bottom (star) and top (solid circle) vacancies extracted from fits in (a). The statistical error due to sample inhomogeneities is on the order of 20\,\% (see Supplemental Information Table~I). The charge state lifetime for Vac\textsubscript{Se} in the lower layer of 2\,ML (turquoise) is plotted next to the 1\,\,ML. (c) DFT calculations of the electron densities of the LUMO of Vac\textsubscript{Se, top} for 1-3\,ML. Left: 
Projections of the LUMO (and LUMO+1 due to neglected spin-orbit coupling) orbital densities $|\Psi_{\text{Vac$_\text{Se}$}}|^2$ on the $z$ axis; right: Side-view of the Vac\textsubscript{Se,top} LUMO isosurfaces in 1-3\,ML.}
\end{figure}

In total, we investigated 22 top and bottom Se vacancies in one to four monolayer WSe\textsubscript{2}. Individual images, spectra, and approach curves are detailed in Fig.~S1 and Fig.~S2. A comparison of the LUMO approach curves for different layer counts is presented in \figref{fig:fig4}(a). We observe a decrease of the saturation current with increasing layer thickness, indicative of an increase in the charge state lifetime. 
\figref{fig:fig4}(b) illustrates the charge state lifetime of top and bottom Se vacancies in the uppermost WSe$_2$ layer of 1-4\,ML \ce{WSe2}. The average charge state lifetimes of the vacancies range from $0.97\pm0.19$\,ps (1\,ML) over $62\pm14$\,ps (2\,ML) and $1.09\pm0.15$\,ns (3\,ML) to $2.31\pm0.66$\,ns (4\,ML).
The extracted values are statistical averages with relative errors of $\approx20\,\%$ related to sample heterogeneities influencing the defect sampling in different locations (Supplementary Information, Tab.~I). Notably, top and bottom Se vacancies exhibit the same charge state lifetime within the experimental error. In the 2\,ML sample, we identified and measured the charge state lifetime also for top and bottom Vac\textsubscript{Se} in the lower layer (turquoise), featuring much higher saturation currents that are comparable with those found in the 1\,ML sample. This confirms our previous assumption that the charge transfer rate from defect orbital to the substrate indeed predominantly depends on their separation. Our values for 1\,ML are also consistent with charge separation dynamics seen in similar WS\textsubscript{2}/graphene heterostructures~\cite{krause_Microscopic_2021}. While the vacancy position within a layer has negligible impact on the charge state lifetime, every spacer layer strongly decouples the defect orbitals causing an increase of the charge transfer time by more than three orders of magnitude from one to four layers.

A strong increase of the charge state lifetime in \figref{fig:fig4}(b) is evident, however the correlation follows a sub-exponential trend, contrasting the expected scaling of quantum tunneling with increasing barrier width. This sub-exponential trend can be explained by the increasing defect orbital delocalization with the layer count. 
Indeed the calculated Vac$_\text{Se,top}$ defect orbitals in 2\,ML and 3\,ML WSe$_2$ is partially delocalized in the lower-lying layers as seen in \figref{fig:fig4}(c).

\section{Summary}
In summary, we have have investigated the layer-dependent binding energy and charge state lifetime of in-gap defect states of individual Se vacancies in WSe\textsubscript{2}. We observe a transition in the Vac\textsubscript{Se,top} LUMO binding energy from a deep center ($E_\text{b} = 594$\,meV) in monolayer (1\,ML) WSe\textsubscript{2} to a shallow defect ($E_\text{b} = 82$\,meV) in 4\,ML WSe$_2$, concurrent with a decrease in the quasiparticle band gap. The average charge state lifetime of Vac$_\text{Se}$ defect states is determined from the saturation of the tunneling current in STM approach curves. We observe a significant increase in the charge lifetime from $0.97\pm0.19$\,ps in 1\,ML to $2.31\pm0.66$\,ns in 4\,ML WSe$_2$, primarily attributed to the distance-dependent suppression of resonant tunneling of the defect-bound charge carrier into the graphene substrate. The sub-exponential growth of the charge lifetime with layer thickness suggests an expansion (delocalization) of the defect wavefunction due to enhanced dielectric screening of the surrounding layers supported by DFT calculations [\figref{fig:fig4}(c)]. Furthermore, we systematically analyze d$I$/d$V$ spectra as a function of tip--sample distance, accessing the exponential and saturated tunneling regime of sequential electron tunneling. Saturation effects are modeled using a two-step transfer rate equation, complemented by DFT calculations of the charge distribution. These findings provide a comprehensive picture of the layer-dependent binding energy and inter-layer charge transfer process between deep levels in TMDs on a graphene substrate.


\section*{References}
\bibliography{references.bib}


\clearpage
\section*{Figures}

\section*{Data availability}
The data that support the findings of this study are available from the corresponding author upon reasonable request.

\section*{Acknowledgements}
This research was supported by the Swiss National Science Foundation under Grant No. 200020-212875 and by the NCCR MARVEL, a National Centre of Competence in Research, funded by the Swiss National Science Foundation (grant number 205602). We also greatly appreciate financial support from the Werner Siemens Foundation (CarboQuant).
We thank Oliver Gröning and Spencer Eve Ammerman for fruitful discussions as well as Pascal Ruffieux and Roman Fasel for general support. L.B., J.A. and B.S. appreciate funding from the European Research Council (ERC) under the European Union’s Horizon 2020 research and innovation program (Grant agreement No. 948243). 
D.C.F., C.D., and J.A.R. acknowledge funding from the Center for Atomically Thin Multifunc-tional Coatings (ATOMIC), an NSF-IUCRC, under award \#2113864, and through NSF-ECCS Award 2202280. R.T., C.D., and J.A.R. acknowledge funding from the 2D Crystal Consortium, National Science Foundation Materials Innovation Platform, under cooperative agreement DMR-1539916. DFT calculations were performed at the Swiss National Supercomputing Centre (CSCS) under project ID s1295. 
For the purpose of Open Access, the author has applied a CC BY public copyright license to any Author Accepted Manuscript version arising from this submission.

\section*{Author contributions}
\textsuperscript{\textdagger} These authors contribute equally to this work.

\section*{Competing interests}
The authors declare no competing interests.

\section*{Supporting information}
\textbf{Supplementary information} The online version contains supplementary material available at https://doi.org/ \\

\end{document}


\title{Supplemental Material:\texorpdfstring{\\}{}Layer-Dependent Charge State Lifetime of Single Se Vacancies in \ce{WSe2}} 
\author{Laric Bobzien\,\orcidlink{0000-0000-0000-0000} \textsuperscript{\textdagger}}

\affiliation{nanotech@surfaces Laboratory, Empa -- Swiss Federal Laboratories for Materials Science and Technology, D\"ubendorf 8600, Switzerland}

\author{Jonas Allerbeck\,\orcidlink{0000-0002-3912-3265} \textsuperscript{\textdagger}}
\affiliation{nanotech@surfaces Laboratory, Empa -- Swiss Federal Laboratories for Materials Science and Technology, D\"ubendorf 8600, Switzerland}

\author{Nils Krane\, \orcidlink{0009-0004-6970-0846}}
\affiliation{nanotech@surfaces Laboratory, Empa -- Swiss Federal Laboratories for Materials Science and Technology, D\"ubendorf 8600, Switzerland}

\author{Andres Ortega-Guerrero\,\orcidlink{0000-0002-0065-0623}}
\affiliation{nanotech@surfaces Laboratory, Empa -- Swiss Federal Laboratories for Materials Science and Technology, D\"ubendorf 8600, Switzerland}

\author{Zihao Wang}
\affiliation{Department of Materials Science and Engineering, The Pennsylvania State University, University Park, PA 16082, USA}

\author{Daniel E. Cintron Figueroa}
\affiliation{Department of Materials Science and Engineering, The Pennsylvania State University, University Park, PA 16082, USA}

\author{Chengye Dong}
\affiliation{Two-Dimensional Crystal Consortium, The Pennsylvania State University, University Park, PA 16802, USA}


\author{Carlo A. Pignedoli\,\orcidlink{0000-0002-8273-6390}}
\affiliation{nanotech@surfaces Laboratory, Empa -- Swiss Federal Laboratories for Materials Science and Technology, D\"ubendorf 8600, Switzerland}

\author{Joshua A. Robinson\,\orcidlink{0000-0002-1513-7187}}
\affiliation{Department of Materials Science and Engineering, The Pennsylvania State University, University Park, PA 16082, USA}
\affiliation{Two-Dimensional Crystal Consortium, The Pennsylvania State University, University Park, PA 16802, USA}
\affiliation{Department of Chemistry and Department of Physics, The Pennsylvania State University, University Park, PA, 16802, USA}

\author{Bruno Schuler\,\orcidlink{0000-0002-9641-0340}}
\email[]{bruno.schuler@empa.ch}
\affiliation{nanotech@surfaces Laboratory, Empa -- Swiss Federal Laboratories for Materials Science and Technology, D\"ubendorf 8600, Switzerland}

\date{\today}        
\pacs{}
\maketitle

\tableofcontents
\newpage
\clearpage

\section{Methods}

\subsection{Sample preparation}

\begin{figure}[b!]
\centering
\includegraphics[width=\textwidth]{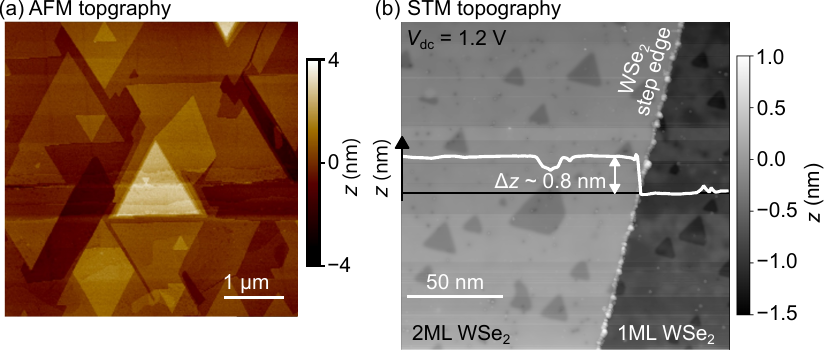}
\caption{Microscopic and nanoscopic overview of sample. (a) AFM image of a 5\,\si{\micro \meter} $\times$ 5\,\si{\micro \meter} area showing triangular islands with different thicknesses of TMD layers. The directionality of the triangle relates to the stacking order of WSe\textsubscript{2} layers. (b) STM topography showing a step edge from 2\,ML (left) to 1\,ML (right). Dark triangles correspond to different thickness of the few-layer graphene substrate, which is less dominant in other sample regions. The effective height difference $\Delta z \approx 0.8$\,nm, corresponds to a convolution of topographic and electronic contrast at the bias voltage $V_\mathrm{dc}$.
\label{fig:SI_fig1}}
\end{figure}

\ce{WSe2} samples are grown on conductive silicon carbide (c-SiC) with few-layer epitaxial graphene via metal organic chemical vapor deposition MOCVD. Intercalation with hydrogen terminates the dangling silicon bonds providing homogeneous electrostatic background. \wse\ synthesis is achieved in a custom-built Kurt J. Lesker vertical cold wall reactor chamber. The QFEG/c-SiC is loaded on an inductively heated SiC coated graphite susceptor, the chamber pressure is then raised to 600\,Torr in a hydrogen environment. Initially, a 10\,min anneal is performed at 350\,°C to remove any volatile surface bound species. Afterwards, the temperature is increased to 800\,°C at a ramp rate of 50°C/min. Once the temperature reaches 800\,°C, both W(CO)\textsubscript{6} and H\textsubscript{2}Se reactants are introduced to the growth chamber for 30\,min at controlled flow rates of 2.6$\cdot10^{-3}$ standard cubic centimeters per minute (sccm) and 78\,sccm, respectively. Then, W(CO)\textsubscript{6} flow is shut off while hydrogen selenide flow continues at 78\,sccm for another 10\,min. Finally, the chamber is cooled down at a rate of 30\,°C/min and H\textsubscript{2}Se flow is shut off at 300\,°C. A representative $5\times$5\,µm\textsuperscript{2} atomic force microscopy image is taken with a Brucker Dimension Icon AFM, shown in \figref{fig:SI_fig1}(b) along with a STM image across a step edge \figref{fig:SI_fig1}(c).

\subsection{Definition of z\textsubscript{0} set point and choice of V\textsubscript{dc}}
Significant orbital contrast and reproducibility across samples and different layer thicknesses requires a definition for the default tunnel set point $z_0 = z (V_0, I_0)$. Here, we define this set point approximately 200\,mV above the dominant onset of the conduction band (CB) of pristine WSe\textsubscript{2}. For multilayer samples, CB states become available at reduced voltages, such that the voltage set point must be adapted. We use $V_0=1.5$\,V for 1\,ML and  $V_0=1.2$\,V for 2-4\,ML samples, both at $I_0=100$\,pA. Low-lying CB states in 3-4\,ML samples are weak such that the set point requires no further adjustment between 2-4\,ML samples. This definition enables reproducible orbital identification at an approximate tip--sample distance of 6-8\,\si{\angstrom} from the TMD surface, where approach curves can be reliably taken without damaging the tip. For the purpose of absolute voltage calibration as discussed below, we assume an absolute tip--sample distance of $(7\pm 2)$\,\si{\angstrom} for the $z_0$ set points as defined above.

\subsection{Creation of selenium vacancies via sputtering}
Chalcogen vacancies in few-layer TMDs can be generated via annealing or ion bombardment in UHV\cite{ma_Controlled_2013, schuler_Large_2019,trishin_Electronic_2023}. During high-temperature annealing ($\approx$ 600\,°C) vacancies often form close to Mo impurities creating defect complexes. To predominantly create isolated vacancies in the uppermost layer, sputtering is the more suitable method. According to the recipe in reference \cite{ma_Controlled_2013}, we used argon ions for sputtering. At an argon pressure of 5.2$\cdot10^{-6}$\,mbar we sputtered the sample for 1\,second at 0.2\,kV acceleration bias, 0.16\,kV discharge voltage, resulting in sub-\,$\mu$A ion current on the sample. The resulting chalcogen vacancy density is well below 1\,\%.

\subsection{Rate equation model and voltage drop}\label{sec:SI_voltage}
The evolution of \didv(V) spectra with decreasing tip--sample distance was simulated using the rate equation model for a double barrier junction as described in the main text:
\begin{equation}
    \label{eq:SIcurrent}
    I(z) = \frac{e}{\Gamma_\mathrm{S}^{-1} + \Gamma_\mathrm{T}^{-1}e^{2\kappa z}}. 
\end{equation}

To include the bias voltage dependency, we write the hopping rate $\Gamma_T$ between defect state and tip as an integral over all states between the Fermi level and \vdc:
\begin{equation}
    \label{eq:SIgammaT}
    \Gamma_T (V_\mathrm{dc}) = C \int_0^{V_\mathrm{dc}} dE\,\rho(E),
\end{equation}
where $\rho(E)$ is the density of states of the defect state and $C$ a renormalization parameter. For simplicity, we assume that LUMO and LUMO+1 behave as one single state, \textit{i.e.} they can only be singly occupied in this energy range and they feature the same charge state lifetime.

In the following, the experimental \didv spectrum of a Se top vacancy in bilayer \ce{WSe2} taken at $z_0+1$\,Å [cf. Fig.2c] was used as $\rho(E)$ and the parameter $C$ was defined from Eq.~(\ref{eq:SIcurrent}) and the experimentally observed current of $82$\,pA for $V_\mathrm{dc}=1.2$\,V, $\kappa=1.06\,\mathrm{Å}^{-1}$ and $\tau=55.9$\,ps for this individual measurement. 
Additionally, we set $\rho(E)$ to exactly zero for $E<0.658$\,V since the experimental noise enters the simulated spectra exponentially at close tip--sample distances.

\begin{figure}[h!]
\centering
\includegraphics[width=\textwidth]{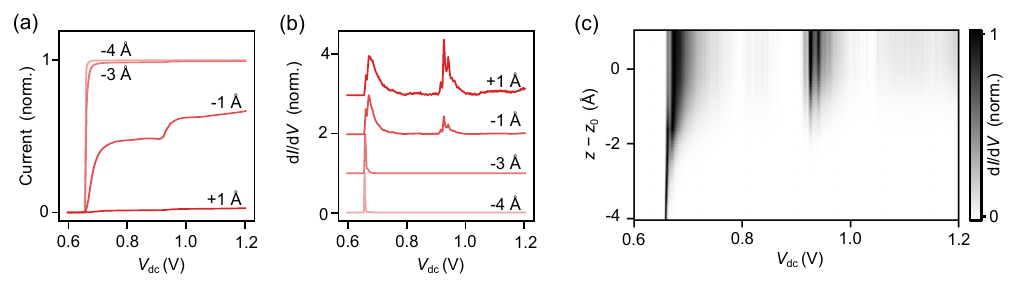}
\caption{\label{fig:SI_fig2}
Calculated evolution of $I_z(V_\mathrm{dc})$ (a) and $\didv_z(V_\mathrm{dc})$ (b) spectra as a function of tip--sample distance $z$ at the $\mathrm{Vac}_\mathrm{Se}$ in bilayer \wse. The calculations are based on the experimental \didv spectrum taken at $z_0+1$\,Å and the effect of a voltage drop over the \wse\ layers is omitted. (c) Normalized \didv(z,\vdc) map, expanding the data in panel (b). Without considering any voltage drop, the peak amplitude in the \didv spectra shifts towards lower bias voltage with decreasing tip--sample distance.
}
\end{figure}

Figure~\ref{fig:SI_fig2}(a) displays the simulated $I_z(V_\mathrm{dc})$ spectra, which were obtained as described above.
At large tip--sample distances, the system is in the exponential tunneling regime ($\Gamma_S \gg \Gamma_T$) and the current increases continuously with increasing bias voltage.
This assumption does not hold anymore when the tip is brought closer to the sample and $\Gamma_T$ becomes equal or larger than $\Gamma_S$, leading to a saturation due to the finite lifetime of the charge state.
Once saturation is reached, the bias modulation by the lock-in amplifier does not cause a change in the current signal and as a consequence the measured \didv signal vanishes [see \figref{fig:SI_fig2}(b)].

Quantitative comparison with the experimental data shows that this simple model does not fully match the intensity evolution of the LUMO+1 resonances. This mismatch is due to the assumption of equal charge-state lifetime for LUMO and LUMO+1. However, the LUMO+1 does have a shorter charge state lifetime, as discussed in the main text, leading to an increase of the saturation current as soon as the tunneling channel via LUMO+1 is possible. Consequently, the LUMO+1 resonances can still be observed at close tip--sample distances, even though the current through the LUMO is already saturated.

For small tip--sample distances, the saturation of the current can already be reached at bias voltages smaller than the first elastic peak of the LUMO, causing an apparent shift to lower bias voltages for the lowest energy peak in the \didv\ spectra, as depicted in Figure~\ref{fig:SI_fig2}(c).
This shift is not visible in the experiment because of a stronger opposing shift towards higher bias voltages, related to the $z$-dependent voltage drop over the \wse\ layers.
In order to reproduce the experimental data, we approximate the effect of the voltage drop via a plate capacitor model:
\begin{equation}
    \label{eq:voltageDrop}
    V_\mathrm{eff.}(\Delta z) = V_\mathrm{dc} \left [ 1+\frac{d}{(z_0+\Delta z)\epsilon_r }\right ], 
\end{equation}
using $z_0=9$\,Å for the distance between tip and defect state, $d=13$\,Å for the thickness of bilayer \wse~\cite{schutte_crystal_1987}, and a dielectric constant of $\epsilon_r=6.5$ for the \wse\ layers~\cite{hou_Quantification_2022}. The resulting spectra are displayed in [Fig. 2d,f] in the main manuscript.

The same voltage correction is applied to calculate binding energy [Fig.~1(d)], spin-orbit coupling [\figref{fig:SI_bandgap}(b)], and band gap energy [\figref{fig:SI_bandgap}(c)], assuming $(7\pm 2)\,\si{\angstrom}$ absolute tip-sample distance. A $\pm 2\,\si{\angstrom}$ uncertainty results in relative errors of $\pm4\,\%$ (1\,ML), $\pm7\,\%$ (2\,ML), $\pm9\,\%$ (3\,ML), and $\pm11\,\%$ (4\,ML).

\subsection{DFT modelling of neutral Vac\textsubscript{Se}}

The calculations were conducted via a Q\textsubscript{UANTUM} ESPRESSO~\cite{Giannozzi_2009,Giannozzi_2017,Giannozzi2020} based AiiDAlab\cite{YAKUTOVICH2021} application\cite{aiidalab-qe-github-repo,UHRIN2021110086}. We considered monolayer to four layer systems, focusing on configurations with pristine structures as well as those with Se vacancies. These analyses were conducted using a supercell approach, specifically employing a supercell dimension of 5 $\times$ 5 $\times$ 1 for each configuration. The Perdew-Burke-Ernzerhof (PBE) \cite{perdew_Generalized_1996} parameterization was used for the exchange correlation functional. Cutoff up 50 (400) Ry was used for the plane waves expansion of the wavefunction (charge density). PAW pseudo-potentials derived from the SSSP library~\cite{SSSP_cite} were used for the representation of the ionic potentials. The Brillouin Zone was sampled using a 6 $\times$ 6 $\times$ 1 mesh grid. Spin-orbit coupling effects were included for the band structure and density of states calculations using the full relativistic pseudo potential from PseudoDojo library~\cite{PseudoDojo_cite} with cutoffs of 100 (400) Ry was used for the plane waves expansion of the wavefunction (charge density). No SOC was included to extract the defect state wavefunctions in Fig.~4.
These calculations were included to the AiiDAlab application using an external plugin~\cite{aiidalab-qe-soc}.
STM images were generated using the Critic2 software~\cite{Critic2_cite}.

\subsubsection{Interactions between graphene and a neutral selenium vacancy in multilayer \ce{WSe2}}

Density functional theory (DFT) calculations were performed to model the interaction between a graphene monolayer and various layers of \ce{WSe2}, including the selenium vacancy (Vac\textsubscript{Se}), using CP2K v2024.1~\cite{kuhne_CP2K_2020}). The PBE exchange-correlation functional with DFT-D3(BJ)~\cite{grimme_Effect_2011} van der Waals corrections was employed. The orbital transformation (OT)~\cite{Weber2008} method was used, with plane-wave and relative energy cutoffs set at 700 Ry and 70 Ry, respectively, and a 5-level multigrid mapping. Goedecker-Teter-Hutter (GTH) pseudopotentials\cite{goedecker1996separable} were used, along with mixed Gaussian and plane-wave basis sets\cite{VandeVondele2007}. The TZV2P MOLOPT basis set described the carbon atoms (TZV2P-MOLOPT-GTH-q4), while the shorter range TZV2P-MOLOPT-SR-GTH basis sets described the tungsten (TZV2P-MOLOPT-SR-GTH-q14) and selenium (TZV2P-MOLOPT-SR-GTH-q6) atoms.

\begin{figure}[t!]
\centering
\includegraphics[width=1.0\textwidth]{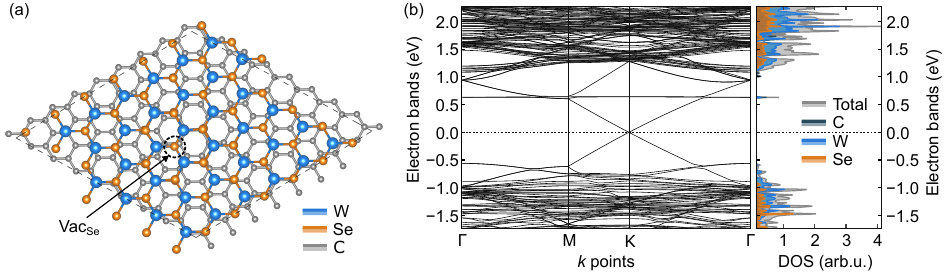}
\caption{(a) Heterostructure of graphene and a \ce{WSe2} monolayer, including a neutral Vac\textsubscript{Se} defect. (b) Band structure calculation and projected density of states (DOS) with individual contributions from different elements.}
\label{fig:gr_tmd_hetero} 
\end{figure}

The graphene and \ce{WSe2} heterostructure was constructed using Hetbuilder~\cite{hetbuilder}. Figure~\ref{fig:gr_tmd_hetero} shows the band structure calculations performed with Q\textsubscript{UANTUM} ESPRESSO. Hybridization between the orbitals of graphene and the defect states occurs at the M point of the hexagonal supercell's Brillouin zone. To model this accurately with the CP2K code, the heterostructure was transformed so that the hybridization occurs at the Gamma point. The rectangular supercell, derived from the hexagonal heterostructure, displays this hybridization at the Gamma point. We used a slab structure built from the rectangular heterostructure for 1, 2 and 3 layers of \ce{WSe2} with a single selenium vacancy.
The cell parameters were A = 34.1608 Å, B = 39.4456 Å, and C = 60 Å. Geometry optimizations of atomic coordinates were performed for all four models.

\subsubsection{Density functional theory including spin orbit coupling in a neutral selenium vacancy in multilayer \ce{WSe2}}

\begin{figure}[h!]
\centering
\includegraphics[width=0.8\textwidth]{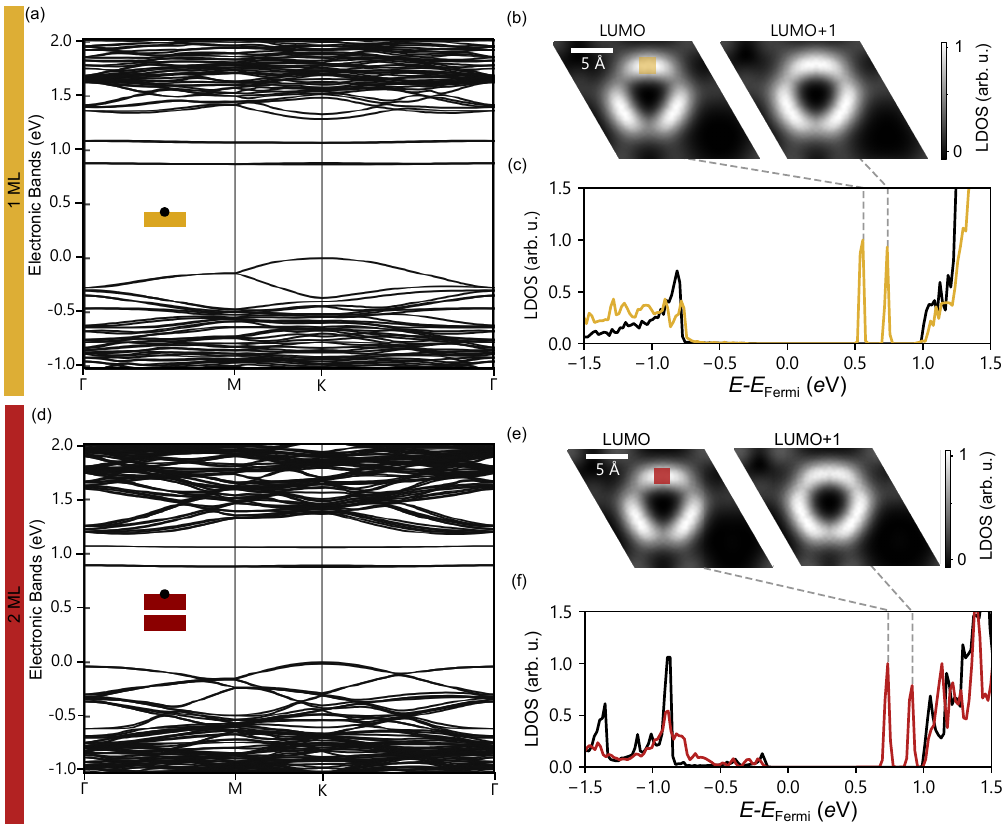}
\caption{\label{fig:SI_DFT}
Density functional theory modelling with spin-orbit coupling of a top Se vacancy in 1ML and 2ML \wse. Electronic band structure for 1ML (a) and 2ML (d). The non-dispersive nature of the in-gap states is well captured for the supercell size of 5$\times$5$\times$1. Images of the local density of state for the LUMO and LUMO+1 for 1ML (b) and for 2ML (e). The LDOS spectrum of the pristine supercell (black) and the vacancy supercell (color) is shown for 1ML (c) and 2ML (f). A smearing of 14\,meV and a sampling box of 2$\times$2$\times$2\,\AA$^3$ (colored squares) at the orbital lobe 7\AA\ above the Se plane was used to calculate the LDOS.
}
\end{figure}

We model Se top vacancies with density function theory (DFT) including spin-orbit interaction for 1ML and 2ML of \ce{WSe2}. In the electronic band structure, the dipserionless spin-orbit split in-gap defect states indicate a sufficiently large supercell size of 5\,x\,5\,x\,1., compare \figref{fig:SI_DFT}  (a) and (d). The maps of the local density of states (LDOS) compare excellently to the experimental equivalents, \figref{fig:SI_DFT}  (b) and (e). Lastly, the LDOS spectrum extracted by cube on the defect lobe and its comparison to the pristine LDOS spectrum compare outstandingly well to the experimentally measured spectra, \figref{fig:SI_DFT}  (e) and (f), neglecting the offset in Fermi energy. In particular, the shape of the valence band in the 2ML case affirms the choice of position and size (2$\times$2$\times$2\,\AA$^3$ (colored squares) at the orbital lobe 7\AA\ above the Se plane) of the cube where we extract the LDOS from.

\subsection{Charge-State Lifetime of the LUMO+1}\label{sec:SI_LUMO+1}

For the LUMO, the saturation current ($I_\text{sat}$) is inversely proportional to the charge state lifetime ($\tau_\text{LUMO}$). However at a voltage resonant with the LUMO+1, tunneling through LUMO and LUMO+1 is possible. Here, we neglect relaxation from LUMO+1 to LUMO, since a spin flip would be required. Then, the saturation current is given by 

\begin{equation}
    \frac{I_\mathrm{sat}}{e} = \frac{1}{1+p} \left( \frac{p}{\tau_\mathrm{LUMO}} + 
    \frac{1}{\tau_\mathrm{LUMO+1}} \right),
    \label{Eq:S4}
\end{equation}

where the charge state life times of both states, $\tau_\mathrm{LUMO}$ and $\tau_\mathrm{LUMO+1}$ contribute. The ratio of tunneling into LUMO and LUMO+1 ($p$) is derived by the energy dependent tunneling probability through a constant potential barrier $p=e^{-2\Delta\kappa z}$, where

\begin{equation}
    \Delta\kappa=\frac{\sqrt{2m_\mathrm{e}}}{\hbar}(\sqrt{\Phi-E_\mathrm{LUMO}}-\sqrt{\Phi-{E_\mathrm{LUMO+1}}}).
\end{equation}

Using $E_\mathrm{LUMO}=1$\,$e$V, $E_\mathrm{LUMO}=1.3$\, $e$V, $\Phi=4.5$\, $e$V and $z=5$\,Å, for a top Se vacancy in 1ML \wse, we obtain $p=0.66$. Therefore, 40\% of the electrons tunnel via the LUMO orbital ($p/(p+1)$), as compared to 60\% via the LUMO+1 orbital ($1/(p+1)$) at the bias voltage resonant with the LUMO+1 state.
By solving Eq.~\ref{Eq:S4} we obtain

\begin{equation}
\label{eq:effectivelifetime}
    \tau_\mathrm{LUMO+1} =  1 / \left( (1+p) \frac{I_\text{sat}}{e}-\frac{p}{\tau_\mathrm{LUMO}}\right) = 0.51 \mathrm{\, ps},
\end{equation}

using $\tau_\mathrm{LUMO}=0.82$\,ps and $e/I_\text{sat}=0.60$\,ps derived from the fits in \figref{fig:SI_overview1}. \\

The reduced charge state lifetime of the LUMO+1 of 0.51\,ps vs 0.82\,ps for the LUMO can be explained by the difference in the orbital-dependent electron density and a reduced tunneling barrier into the substrate for the higher-energy LUMO+1. We can disentangle their relative contribution by modelling the energy-dependent barrier between defect orbital and substrate. Assuming a constant potential barrier, we expect a reduction of $\tau_\mathrm{LUMO+1}= 0.68\, \tau_\mathrm{LUMO}$, for an effective graphene--defect distance of $5$\,Å. The remaining difference between LUMO+1 and LUMO lifetimes we attribute to their slightly different orbital distributions. We estimate this contribution to add another reduction of $\tau_\mathrm{LUMO+1}= 0.9\,\tau_\mathrm{LUMO}$ of the charge state lifetime of the LUMO+1 to explain the $62$\,\% overall reduction.
Similarly, we can extract the charge state lifetime for the LUMO+1 in a vacancy in 2ML, as in [Fig. 2a] in the main text, for $\tau_\mathrm{LUMO}=55.9$\,ps and $I_\text{sat}/e=40.4$\,ps. We predict 
 $\tau_\mathrm{LUMO+1}=0.7\,\tau_\mathrm{LUMO}=39.1$\,ps.
\newpage
\section{Supporting experimental data}

\subsection{Overview of Vac\textsubscript{Se} in 1-4 monolayer WSe\textsubscript{2}}

\begin{figure}[H]
\centering
\includegraphics[width=\textwidth]{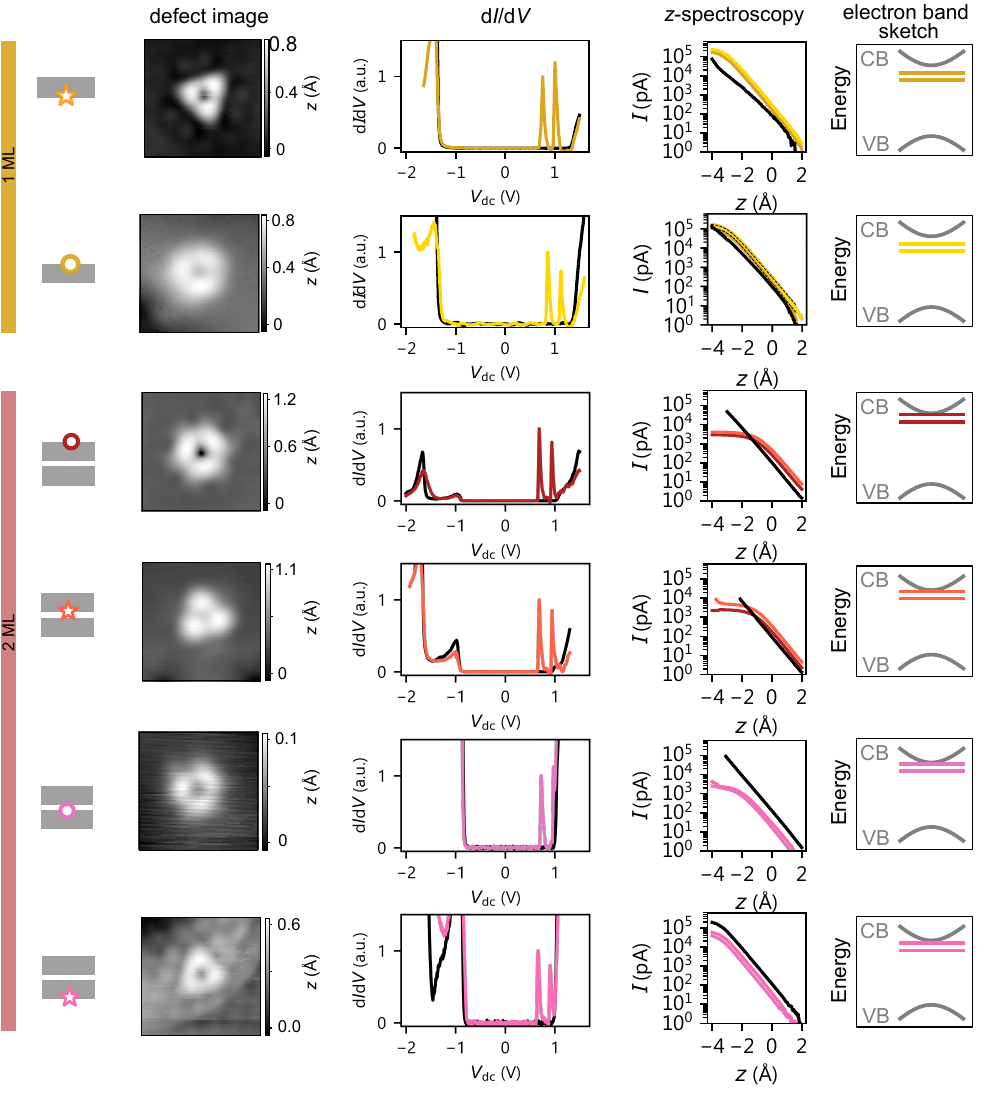}
\caption{\label{fig:SI_overview1}
Defect overview 1/2. Complete characterization of all defects studied in the main manuscript 1-2ML. Columns from left to right: STM topography at ($V_0$,$I_0$), differential conductance (d$I$/d$V$) of the defect orbital (colored) with a reference to pristine WSe\textsubscript{2} (black), and approach curves for the LUMO (dark color), LUMO+1 (light color) and for pristine WSe\textsubscript{2} (black).}
\end{figure}

\begin{figure}[H]
\centering
\includegraphics[width=\textwidth]{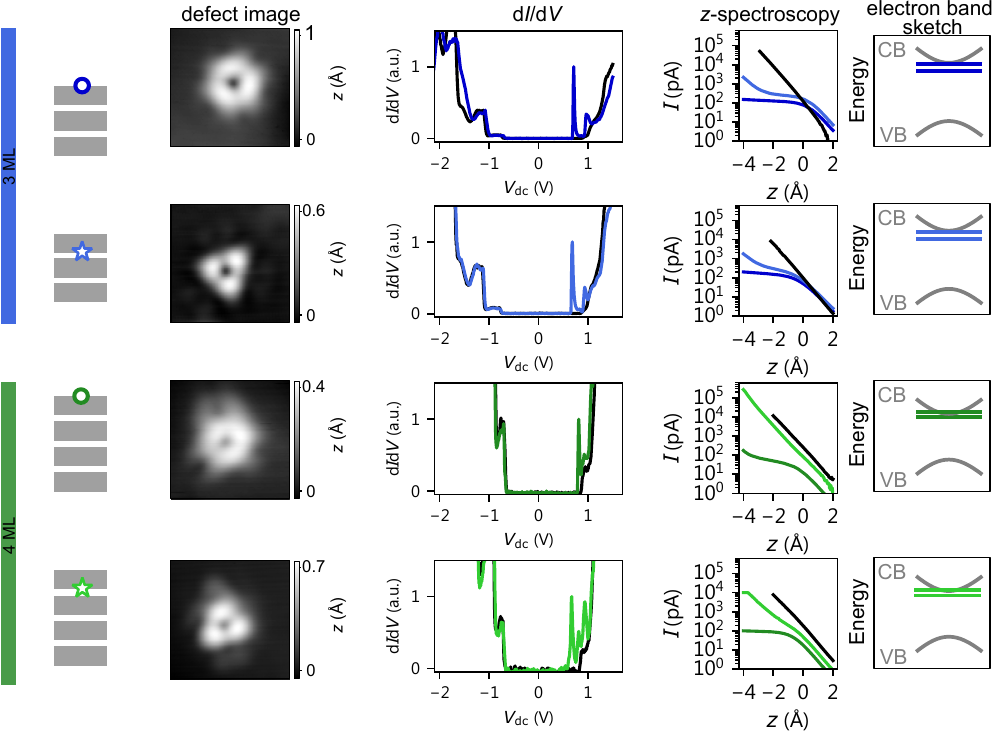}
\caption{\label{fig:SI_overview2}Defect overview 2/2. Complete characterization of all defects studied in the main manuscript 3-4. Columns from left to right: STM topography at ($V_0$,$I_0$), differential conductance (d$I$/d$V$) of the defect orbital (colored) with a reference to pristine WSe\textsubscript{2} (black), and approach curves for the LUMO (dark color), LUMO+1 (light color) and for pristine WSe\textsubscript{2} (black).}
\end{figure}

Figures \ref{fig:SI_overview1} and \ref{fig:SI_overview2} provide an overview of each defect configuration discussed in the manuscript. For each layer thickness ranging from 1ML to 4ML we investigated top and bottom vacancies of the upper layer. Columns of the figure show STM topography, d$I$/d$V$ spectra, tip approach curves obtained at the defect site (bright lobe) referenced to pristine WSe2 (black curves), and a schematic band diagram. Colored approach curves correspond to different bias voltages above LUMO and above LUMO+1. The bias is chosen such that the defect resonance remains below $V_\mathrm{dc}$ for all values of $z$.

\subsection{Statistical analysis of the charge state lifetime}
To support our findings we present a statistical analysis of all defects investigated on the sample in Table\,\ref{table:SI_tranfertime}. The main source of variation are neighboring defects in the same or adjacent layers. While defects in the same layer are usually apparent and can be omitted, they are difficult to resolve in subsurface layers. Most frequently find Mo substitutional defects next to Se vacancies.

\begin{table}[H]
\centering
\begin{tabular}{|l|l|l|l|l|l|}
\hline
\multicolumn{2}{|l|}{\textbf{WSe\textsubscript{2} layers}}             & 1 & 2 & 3  & 4 \\ \hline
\multirow{3}{*}{Vac\textsubscript{Se} LUMO}     & $N$                       & 3 & 10 & 7 & 2 \\ \cline{2-6} 
                                           & $\bar{\tau}$ (ps)         & 0.97 & 62 & 1085 & 2308 \\ \cline{2-6} 
                                           & $\sigma_\tau$ (ps) ($\sigma_\tau/\tau$)    & 0.19 (20\%) & 14 (22\%) & 147 (14\%) & 657 (27\%) \\ \hline

\end{tabular}
\caption{
Experimental statistics of the charge state lifetime at the Vac\textsubscript{Se} LUMO in few-layer WSe\textsubscript{2}. $N$ is the combined number of top and bottom vacancies analyzed at each layer thickness, $\Bar{\tau}$ the mean value, and $\sigma_\tau$ ($\sigma_\tau/\tau$) the absolute error (relative) error.}
\label{table:SI_tranfertime}
\end{table}

\subsection{Additional data for Vac\textsubscript{Se} in 3\,ML WSe\textsubscript{2}}
\begin{figure}[b!]
\includegraphics[width=\textwidth]{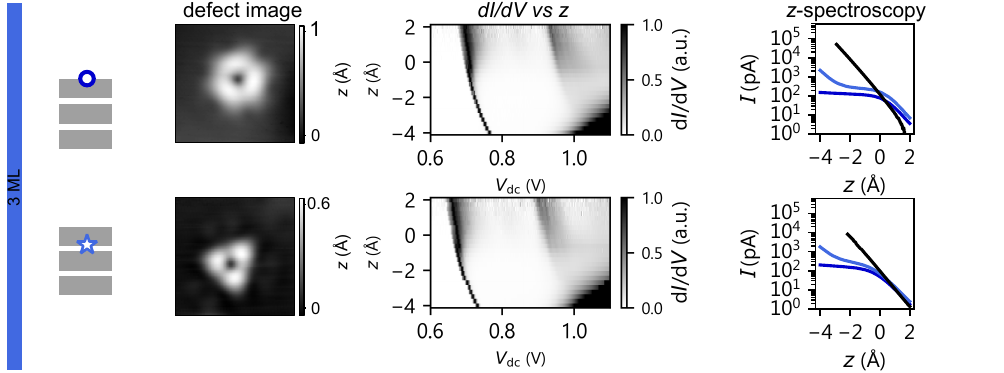}
\caption{\label{fig:SI_3ML}
Top and bottom Se vacancies in 3ML WSe\textsubscript{2}. Left: Constant current STM images at ($I_0=100$\,pA, $V_0=1.2$\,V) of the defect orbitals. Middle: Differential conductance d$I$/d$V$ spectra as a function of tip--sample distance $z$, showing the $z$-dependent shift of the onset of LUMO and LUMO+1 similar to [Fig. 2]. Dark areas of high d$I$/d$V$ in the bottom right corner correspond to tunneling contribution of the CB at small $z$. In the approach curves (right), this contribution of the CB leads to an exponential increase in the current for $z<z_0-2\,\si{\angstrom}$ most notably for the LUMO+1.}
\end{figure}

In addition to [Fig. 3] of the main manuscript, \figref{fig:SI_3ML} shows the experimental \didv-$z$ map that emphasizes a strong overlap of conduction band and LUMO+1 at reduced tip--sample distances. Lower panels show correponding data of a bottom vacancy in the top layer of 3\,ML WSe\textsubscript{2}. 

\subsection{WSe\textsubscript{2} bandgap and Vac\textsubscript{Se} spin-orbit splitting}

\begin{figure}[t!]
\centering
\includegraphics[width=0.6\textwidth]{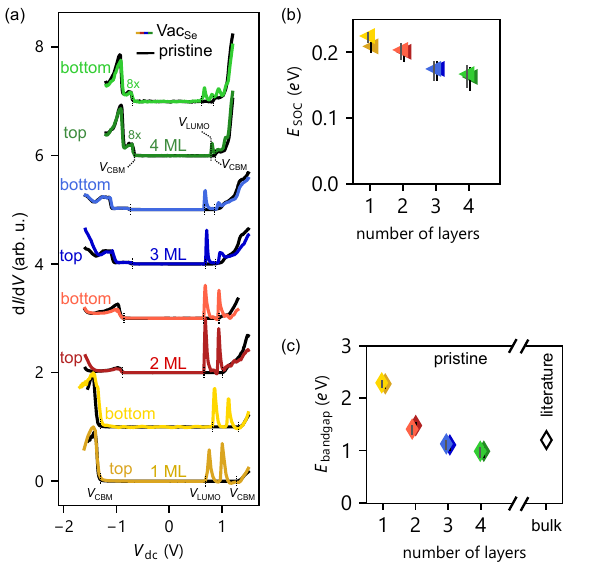}
\caption{\label{fig:SI_bandgap}
Differential conductance spectra of Va\textsubscript{Se} in few-layer WSe\textsubscript{2}. (a) d$I$d$V$ spectra at Vac\textsubscript{Se} (colored) and pristine WSe\textsubscript{2} (black) for all defects in [Fig. 4a]. Dotted ticks mark the onset voltage of LUMO as well as valence band maximum (VBM) and conduction band minimum (CBM). (b) The spin-orbit splitting, as the energy difference between LUMO and LUMO+1 is plotted.
(c) The band gap extracted from (a) for pristine WSe\textsubscript{2} as a function of layer thickness using the same color code as for the defects in (a). The experimental value for the indirect band gap of bulk  WSe\textsubscript{2} is taken from reference \cite{kam_detailed_1982}. Error bars in panels (b) and (c) reflect an uncertainty of the absolute tip--sample distance $(7\pm 2)\,\si{\angstrom}$ when estimating the voltage drop.}
\end{figure}

\figref{fig:SI_bandgap}(a) summarizes the \didv spectra of top and bottom vacancies in 1-4\,ML WSe\textsubscript{2} used for primary analysis and partially presented in [Fig. 1b] of the main manuscript. In addition to the binding energy shown in [Fig. 1d], \figref{fig:SI_bandgap}(b) analyzes the layer-dependent variation of spin-orbit splitting of the Vac\textsubscript{Se} states as seen from the spectra in (a). \figref{fig:SI_bandgap}(b) extracts the fundamental transport bandgap of pristine WSe\textsubscript{2} as measured in the reference spectra. 

\clearpage

\newpage
\bibliography{references}